# Theory of Double Ladder Lumped Circuits With Degenerate Band Edge

Jeff Sloan, Mohamed A. K. Othman, and Filippo Capolino

*Abstract*—A conventional periodic LC ladder circuit forms a transmission line that has a regular band edge between a pass and a stop band. Here for the first time we develop the theory of simple yet unconventional double ladder circuit that exhibits a special degeneracy condition referred to as degenerate band edge (DBE). The degeneracy occurs when four independent eigenstates coalesce into a single eigenstate at the DBE frequency. In addition to possible practical applications, this circuit may provide insight into DBE behavior that is not clear in more complex systems. We show that double ladder resonators exhibit unusual behavior of the loaded quality factor near the DBE leading to a stable resonance frequency against load variations. These two properties in the proposed circuit are superior to the analogous properties in single ladder circuits. Our proposed analysis leads to analytic expressions for all circuit quantities thus providing insight into the very complex behavior near degeneracy points in periodic circuits. Interestingly, here we show for the first time that DBE is obtained with unit cells that are symmetric along the propagation direction. The proposed theory of double ladders presented here has potential applications in filters, couplers, oscillators, and pulse shaping networks.

*Index Terms*—Degenerate Band Edge, Cavity Resonators, Circuit Theory, Slow-Wave Circuits, Ladder Oscillators.

## I. INTRODUCTION

PERIODIC structures and circuits have been utilized in many RF components and devices due to their unique properties such as the existence of electromagnetic band edges and bandgaps [1]–[3]. The "band edge" condition refers to a point in the phase-frequency dispersion relation which separate a pass band and a stop band in a periodic structure. Dispersion diagrams are associated with structures of infinite length with a unit cell periodically repeated. The band edge is accompanied with a significant increase in delay and quality factor of periodic resonators. The band edge is also associated with degenerate eigenstates of the field quantities (electric and magnetic field states in waveguides, or voltage and current states in circuits) and they correspond to a standing resonant mode. Beyond the band edge, a band gap is typically formed in which signal flow is inhibited inside the periodic circuit resulting in only attenuation (evanescent wave in electromagnetic band gap structure).

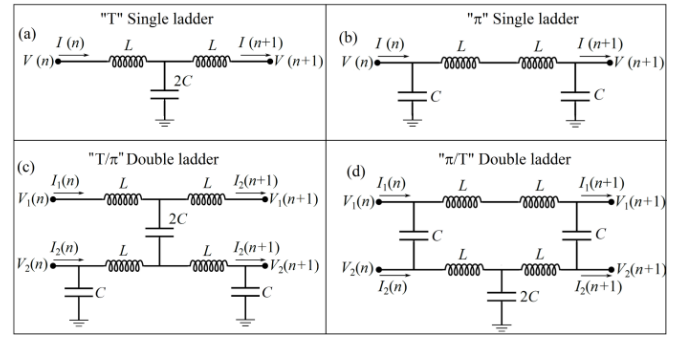

Fig. 1. (a), (b) Unit cells of a conventional periodic single ladder of T and $\pi$ configurations, respectively. (c), (d) Symmetric unit cells with four ports of periodic lumped circuits, called here "*double ladders*", that develop DBE at an angular frequency $\omega_d = 1/\sqrt{LC}$, with T/$\pi$ and $\pi$/T configurations respectively. All voltages are referred to the ground.

A particular class of degeneracy may exist in a periodic circuits where four periodic eigenstates coalesce and form a single degenerate periodic eigenstate [4]–[6]. This condition, explored here, is called a degenerate band edge (DBE), contrary to the scenario in conventional spatially periodic structures where only two periodic eigenstates coalesce forming a regular band edge (RBE). A DBE can be found in periodic structures employing stacks of anisotropic layers [4], [7], [8], structured transmission lines or striplines [9], [10], metallic [11] or optical waveguides [12], [13]. When four periodic eigenstates coincide at the edge of the Brillouin zone, the dispersion relation close to the degeneracy point is characterized by $(\omega_d - \omega) \propto (\varphi - \pi)^4$ where $\omega_d$ is the angular frequency at which this fourth order degeneracy occurs, and $\varphi$ is the state phase delay of circuit quantities from one unit cell to the next. The exponent indicates that this degeneracy condition is of order four. The degeneracy condition in such a class of periodic waveguiding structures [7], [9], [11] is associated with a dramatic reduction of group velocity and an increase in loaded quality factor that is crucial for various application including filters, oscillators, and pulse forming networks for high speed communication [14]–[18]. DBEs were also investigated for directive antenna applications [10], [19]. Slow-wave structures (SWSs) with DBEs, in which phase velocity is much less than the speed of light, allow for superior electron beam synchronism condition that leads to high gain [6] compared to conventional SWSs. It has been also shown that utilizing a DBE in an active devices will lower the

This material is based upon work supported by the Air Force Office of Scientific Research under award number FA9550-15-1-0280 and under the Multidisciplinary University Research Initiative award number FA9550-12-1-0489 administered through the University of New Mexico.
The Authors are with the Department of Electrical Engineering and Computer Science, University of California, Irvine, CA 92697 USA. (e-mail: sloanj@uci.edu, mothman@uci.edu, f.capolino@uci.edu)

oscillation threshold in cavities [20] compared to operating near an RBE [21], in which the latter was shown to have modal oscillation instabilities and mode jumping issues [22], [23].

Here we propose for the first time a periodic lumped double ladder circuit, whose unit cell is made of just a few reactive elements, that develops a degenerate band edge condition (in contrast to previous investigations that were focused on transmission lines).

Because this lumped circuit is very simple, we show for the first time exact analytic steady state solutions for periodic voltage/current eigenstates in periodic double ladders; and the occurrence of an unusual resonance in double ladders with finite size. Our analysis provides a concrete insight into general DBE manifestation and characteristics of ladder lumped circuits for which analytical investigation of resonance and loading effects has not been reported before. It is important to point out that we do not just propose a double ladder filter design in this paper, but also we develop a novel theory of a class of circuits exhibiting a fourth order degeneracy; the applications of which may range from oscillators, pulse compressors and distributed amplifiers.

A degeneracy of order four requires that the unit cell of the periodic structure consists of a four-port network with properly coupled and tuned reactive components. Higher order degeneracies can be achieved with more complicated unit cells with a higher number of ports, yet the same approach detailed in this paper can be utilized. We stress that the proposed circuit is one-dimensional. Two-dimensional [23], and even three-dimensional ladder circuits may also be conceived whose analysis can be carried using the same Bloch-Floquet type solution for 2D and 3D periodic electromagnetic systems [2], but they are outside the scope of this study. The delay characteristics of these ladder circuit are analogous to wave propagation in one-dimensional crystals [24], [25], yet we show the characteristics of state degeneracies in ladder circuits that are analyzed here for the first time.

In this paper, we demonstrate various new concepts and detailed analyses relative to degeneracies in circuits as follows: (ii) we propose a periodic circuit with the minimum number of reactive elements that exhibits a DBE and (ii) we develop analytical theory of such double ladders and explore their unusual characteristics in Sections II through V. (iii) We show that such DBE properties can be obtained in symmetric circuits in Section IV. (iv) We investigate the properties of double ladder resonators made of a cascade of a finite number of cells, and we provide analytic expressions for their design in Section VI. (v) We analyze how the quality factor is affected by loading and importantly how it scales with structure size along with various loading effects. (vi) We show the superior performance of double ladder resonator compared to a single LC ladder with an RBE in Fig. 1, in terms of stability of resonance frequency with loading effects and the anomalous scaling of the quality factor in Section V.

Throughout this paper we assume steady-state monochromatic signals, and phasors are based on the $e^{j\omega t}$ time convention that is implicitly assumed.

## II. DOUBLE LADDER CIRCUIT WITH SYMMETRIC UNIT CELL

An example of a periodic circuit's constitutive symmetric unit cell that develops a DBE in the steady state regime is depicted in Fig. 1(c) and (d). There are other circuits that have DBEs but this circuit is the simplest, in part due to symmetry. It is important to distinguish between two kinds of symmetries. The first is the symmetry of the top and bottom ladders (in double ladder case). There exists a strong asymmetry in the sense that the top ladder does not have nodes with capacitances to the ground, unlike its bottom counterpart. Therefore, the anisotropy here is created (top and bottom ladders are not identical) and it is crucial for achieving the DBE [4], [9], [11]. The second, and the more interesting kind of symmetry is manifested about a cut in the middle of the unit cell (left and right symmetry) which represent symmetry about a perpendicular axis to the signal propagation direction. It is also possible to *cut* this circuit in a variety of ways that yield a variety of unit cells, but the ones that are shown are the simplest due to symmetry. Contrary to most DBE implementations in waveguiding systems, a symmetric unit cell in the wave-propagation direction has not been reported before, see for example in [9], [13], [26]. Moreover, there have been some persistent difficulties in designing circuits with a DBE. One difficulty is that multiple parameters must be simultaneously tuned to get close to a DBE condition with negligible analytical insight. A related issue is that the DBE condition is never met exactly, but only in an approximate way. This can cause problems in numerical calculations; for instance when inverting a matrix. Another difficulty is that the eigenvectors are generally not easy to calculate in a simple way, potentially obscuring some useful information. In fact, our analysis shows for the first time a *mathematically exact* condition in circuits to observe a DBE. Moreover, we believe that the circuit investigated in this paper is the simplest circuit (and one of the simplest physical systems) to exhibit a DBE. The eigenstate description as well as the resonance conditions are simple enough to be put in a very compact and concise analytical form near the DBE. Here we study a symmetric unit cell, which can be cut in either the T/$\pi$ or the $\pi$/T configuration shown in Fig. 1. This portrays also how a double ladder that develops a DBE can be constructed from a simple single LC ladder connected in tandem, using either the conventional $\pi$ or T topologies [15]. We assume lossless, linear reactive components for simplicity in our analysis. Losses in the resonator will be incorporated in Section VI in the form of external loading.

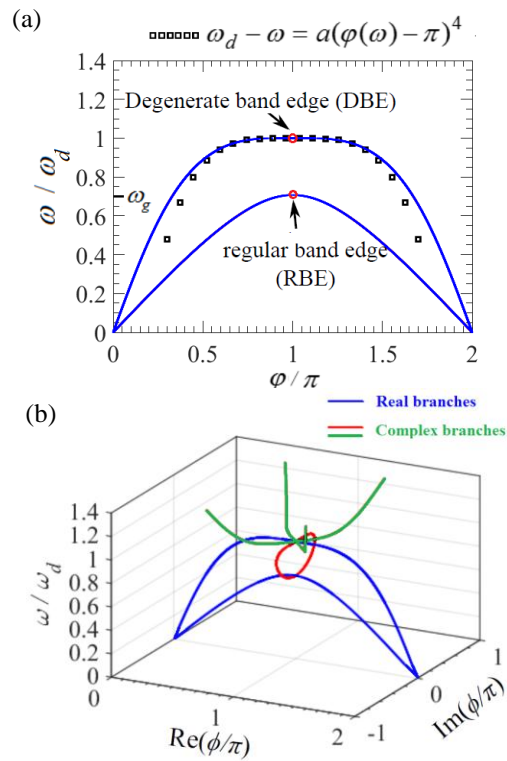

Fig. 2. Dispersion diagram, frequency-phase shift, relative to the unit cell of the periodic, infinitely long, double ladder circuit in Fig.1(c) or (d). Each of the 2 curves corresponds to two (backward and forward) periodic eigenstates. Four eigenstates coalesce at the degenerate band edge (DBE) at $\omega = \omega_d$. Two eigenstates coalesce at the RBE at $\omega = \omega_g$. Square symbols denotes a fitting of the DBE flat feature with $\omega_d - \omega = a(\pi - \varphi(\omega))^4$, and $a$ is the fitting constant $a = \omega_d / 32$. (b) Complex dispersion curve showing complex branches of $\varphi(\omega)$.

*A. Dispersion relation and state vector*

First, we qualitatively explore the voltage/current periodic eigenstates (in a steady state regime) of the infinitely-long periodic circuit whose unit cell is depicted in Fig. 1, and we resort to a rigorous analytical description detailed in Section IV. It is convenient to define a four-dimensional state vector that comprises the phasors for voltages and currents at the $n^{\text{th}}$ unit cell's two ports as seen in Fig. 1 viz.

$$\mathbf{\Psi}(n) = \begin{bmatrix} V_1(n) & V_2(n) & I_1(n) & I_2(n) \end{bmatrix}^T \quad (1)$$

In an infinitely long periodic structure, steady state time-harmonic solutions have the form of periodic eigenstates, i.e., eigenstates where the current and voltage at location $n$ in a unit cell of the infinite periodic circuit in Fig. 1 is a complex multiple of the current and voltage at the same locations in an adjacent unit cell (location $n+1$), as described in the following. We seek those periodic solutions (more properly, pseudo periodic, because they are periodic in space except for the exponential term $e^{-j\varphi(\omega)}$) for the state vectors, where $\mathbf{\Psi}(n)$ is translated to that of the next unit cell $\mathbf{\Psi}(n+1)$ via

$$\mathbf{\Psi}(n+1) = \mathbf{\Psi}(n) e^{-j\varphi(\omega)} \quad (2)$$

for any $n$. Note that here the exponential term $\varphi(\omega)$ can be complex. A purely real $\varphi(\omega)$ implies simply a phase shift from cell to cell, whereas a complex $\varphi(\omega)$ also implies exponential attenuation or growth and must be accounted for completeness. The evolution of this four-dimensional state vector $\mathbf{\Psi}(n)$ from cell to cell is described by a 4×4 transfer matrix $\underline{\mathbf{T}}$ that relates voltages and currents between contiguous cells at location $n$ and $n+1$

$$\mathbf{\Psi}(n+1) = \underline{\mathbf{T}} \mathbf{\Psi}(n) \quad (3)$$

Here and in the following bold fonts indicate vectors whereas underlined bold fonts indicate matrices. The determination of the transfer matrix $\underline{\mathbf{T}}$ for the circuit in Fig. 1 is reported in the Appendix. In general, for an infinite long periodic circuit whose unit cell has four ports as shown in Fig. 1, there are four distinct and independent solutions for the cell-to-cell phase progression at each frequency [see (2)], with $m = 1, 2, 3, 4$, each of which is associated with a periodic eigenstate voltage/current eigenvector $\mathbf{\Psi}_m$. Again, here $\varphi(\omega)$ is allowed to be complex so that it can describe also stop bands. To retrieve periodic solutions for $\mathbf{\Psi}(n)$, i.e., in the form of (2), we solve the eigenvalue problem obtained by combining (2) and (3)

$$\left[ \underline{\mathbf{T}} - e^{-j\varphi(\omega)} \underline{\mathbf{1}} \right] \mathbf{\Psi}(n) = 0 \quad (4)$$

Here $\underline{\mathbf{1}}$ is the 4×4 identity matrix and $\varphi$ is the Bloch phase shift between two adjacent cells. For the circuit in Fig. 1, the four $\varphi_m(\omega)$ solutions are shown in Fig. 2, for $m = 1,2,3,4$, with the circuit parameters given in the Appendix. This is because the number of possible states is twice the number of nodes (excluding ground), or ports, shared by two contiguous unit cells [27], [28]. Analogously, triple ladders circuits are constructed by 6×6 matrices, and so forth. Analysis of higher order ladders can be done by increasing the dimensionality of the $\underline{\mathbf{T}}$ matrix; which is left for another investigation.

Phase progression of an eigenstate, or evanescence (i.e., exponential growth or decay) from cell to cell in the lossless circuit can be understood by examining both the real and imaginary parts of $\varphi_m(\omega)$. For instance, in a lossless structure, an eigenstate whose voltages and currents exhibit only phase progression (propagating in the context of electromagnetic structures) has $\text{Im}[\varphi(\omega)] = 0$, however an evanescent eigenstate (in the stop band of the periodic circuit) has $\text{Im}[\varphi(\omega)] \neq 0$ while $\text{Re}[\varphi(\omega)]$ can be non-zero. We focus on the simple unit cell in Fig. 1 whose dispersion diagram is depicted in Fig. 2 considering the circuit parameters given in the Appendix. The circuit is linear and reciprocal, therefore if $\varphi(\omega)$ is a solution (even complex) then also $-\varphi(\omega)$ is. Furthermore, the dispersion diagram of the periodic structure is periodic, and exhibits a mirror symmetry around the band edge (between pass and stop bands) that occurs at $\varphi = \pi$. A typical dispersion diagram is conventionally illustrated by the real eigenstates *only*, whose phase $\varphi(\omega)$ is purely real, as shown in Fig. 2. A complex representation of the dispersion diagram is also shown in Fig. 2 and can be found in [6], [12], [29] for structures that involve wave propagation.

At low frequency, such that $\omega < \omega_g$, the $\varphi(\omega) - \omega$ relation shows purely real modal phase shifts $\varphi(\omega)$, from cell to cell as in (2), versus frequency as depicted in the dispersion diagram in Fig. 2. At high frequency, such that $\omega > \omega_d$, the circuit exhibits a cutoff where energy flows in the circuit is highly suppressed resembling the high frequency response of a low-pass filter. At an intermediate angular frequency (shown in Fig. 2) we observe two important distinct features of the dispersion diagram.

*B. Band edges and band gaps*

The dispersion diagram (Fig. 2) shows that there are 4 independent periodic eigenstates (corresponding to 4 distinct eigenvalues of the periodic circuit) everywhere except at two specific frequencies, denoting the "band edges". The band edge is a transition condition from power-carrying eigenstates(s) to evanescent eigenstates(s). The band edge point itself represents a standing wave eigenstate that does not allow energy to flow in the periodic circuit (a point of singularity in group delay). There are two band edge conditions:

*1) Regular band edge:* At $\omega = \omega_g$ an RBE is manifested where two branches (phase eigenstates with opposite directions of energy flow) coalesce. Therefore, an RBE is a point of degeneracy of two states' eigenvalues and eigenvectors. Near $\omega = \omega_g$, in virtue of small frequency detuning, the phase-frequency asymptotically behaves as $\omega_g - \omega \propto (\varphi(\omega) - \pi)^2$ with a proportionality constant specific to the circuit and proportional to the non-zero first derivative of the group delay. (For the circuit in Fig. 1, it happens that $\omega_g = \omega_d/\sqrt{2}$). Exactly at the RBE there are three *independent* periodic eigenstates with three regular eigenvectors, which are found by solving the periodic system (4). One of the periodic eigenstates represents a standing resonant mode with infinite group delay. In fact, the RBE designates a transition from two states with phase progression $\omega < \omega_g$ into two purely evanescent states for $\omega > \omega_g$, resulting in a second order degeneracy at $\omega = \omega_g$. The other two eigenstates (the two upper branches) at $\omega = \omega_g$ are independent phase-propagating solutions and are not affected by the RBE. As such, there must exist one degenerate eigenstate in order to provide a complete basis of four eigenvectors and it is constructed with a non-periodic eigenstate (or pseudo-periodic eigenstate described by generalized eigenvector [4], [30], [31]) that grows linearly along the double ladder.

*2) Degenerate band edge:* At $\omega = \omega_d$ a degenerate band edge (DBE) is manifested. A DBE is a point of degeneracy of four states' eigenvalues and eigenvectors (4). The phase-frequency relation near DBE is $\omega_d - \omega \approx a(\pi - \varphi(\omega))^4$ with a proportionality constant specific to the circuit, that provides the non-zero third derivative of the group delay, and this constant will be found in the subsequent analysis. Despite at other frequencies there are generally four independent eigenstates, at the DBE there is only a single, degenerate, periodic eigenstate at $\omega = \omega_d$ comprising a standing resonant mode that is a transition from two states with phase progression and two evanescent ones, for $\omega < \omega_d$, into four purely evanescent states for $\omega > \omega_d$. Since there is only one independent periodic eigenstate solution at $\omega_d$ there must be three non-periodic eigenstates i.e., three pseudo-periodic solutions growing as $n$, $n^2$ and $n^3$ [32], [33] besides phase factor $\exp(-jn\pi)$, where $n$ is the integer index of the unit cell. Therefore, exactly at the DBE, state vectors propagate from cell to cell as $\Psi(n) \propto n^q \exp(-jn\pi)$, with $q = 0, 1, 2, 3$, (see analogues non-periodic solutions in stack of anisotropic layers at the DBE in [32]). The rest of this paper will be dedicated to developing analytic framework for the intriguing properties associated with DBE in the infinitely periodic double ladder first, and then, most importantly, for double ladders resonators with finite size.

## III. CIRCUIT ANALYSIS NEAR THE DBE

Kirchhoff's voltage law (KVL) provides a straightforward route to analyzing steady state solutions in the periodic lumped circuit. It is convenient to define loop currents (phasors) as shown in Fig. 3 for the periodic double ladder with the unit cell as in Fig. 1. We consider the two current loops per unit cell with current phasors $I_n$ and $I'_n$. Current in adjacent unit cells vary by the constant $e^{-j\varphi}$ based on (2) and the band edge condition $\varphi(\omega) = \pi$ corresponds to standing eigenstates. In the following we analyze the state vector (1) in close vicinity of (and not exactly at) the DBE where there are four periodic solutions as in (2). As such, we define an incremental phase angle that is small in magnitude near the DBE. The DBE angular frequency is set to be $\omega_d = 1/\sqrt{LC}$, according to the parameters of the circuit in Fig. 1, with a *characteristic impedance* parameter defined as $Z_c = \sqrt{L/C}$. (This will be evident in Section IV.) Therefore, the relation of loop currents between adjacent cells is $I_{n+1} = I_n e^{-j\varphi(\omega)} = -I_n e^{j\delta(\omega)}$ and we assume that $I'_n = \alpha I_n$ where $\alpha$ is a frequency dependent constant, evaluated as follows.

Application of KVL yields the loop voltage drops

$$I'_n\left[4\omega L - \frac{1}{\omega C}\right] + \frac{1}{2\omega C}(I'_{n+1} + I'_{n-1}) - \omega L(I_n + I_{n+1}) = 0$$
$$I_n\left[2\omega L - \frac{1}{\omega C}\right] + \frac{1}{2\omega C}(I_{n+1} + I_{n-1}) - \omega L(I'_{n-1} + I'_n) = 0 \quad (5)$$

which in turn leads to the two following relations

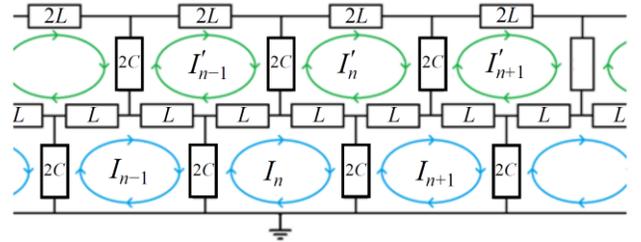

Fig. 3. Loop currents used in the KVL analysis for a double ladder circuit. Near the DBE, the current flowing in the lower half of the circuit represented by the blue loops is much larger in magnitude compared to that flowing in the upper half of the circuit represented by the green loops. Exactly at the DBE only the lower blue loops carry current.

$$\left(\frac{\omega}{\omega_d}\right)^2 \left[\alpha e^{-j\delta(\omega)} - \alpha + 2\right] = \left[1 + \cos(\delta(\omega))\right],$$

$$\left(\frac{\omega}{\omega_d}\right)^2 \left[\left(\frac{e^{j\delta(\omega)} - 1}{\alpha}\right) + 4\right] = \left[1 + \cos(\delta(\omega))\right] \quad (6)$$

By dividing the two equations in (6) we get

$$-\left(1 - e^{-j\delta(\omega)}\right)\alpha^2 - 2\alpha + \left(1 - e^{j\delta(\omega)}\right) = 0 \quad (7)$$

whose solutions for $\alpha$ can be easily written as

$$\alpha = -j\frac{1 \pm \sqrt{1 + 4\sin^2\left(\frac{1}{2}\delta(\omega)\right)}}{2\sin\left(\frac{1}{2}\delta(\omega)\right)} e^{j\frac{\delta(\omega)}{2}} \quad (8)$$

Equations (7) or (8) provide the exact relationship between the phase delay between contiguous unit cells and the ratio $\alpha$ of the upper and lower current loops within a unit cell. We show now a convenient approximation to derive both the phase delay and the current ratio as a function of angular frequency near the DBE. Near the DBE $|\delta| \ll 1$ and we rewrite (8) using the first order Taylor expansion for small $\delta$ as

$$\alpha \approx j\delta(\omega)e^{-j\frac{\delta(\omega)}{2}} \tag{9}$$

Then, substitution of $\alpha$ into (6), and keeping only terms with lowest power $\delta(\omega)$ leads to

$$\frac{\omega}{\omega_d} \approx \left(1 - \delta^4/32\right) \tag{10}$$

that yields to a simple form of the asymptotic dispersion relation

$$\omega_d - \omega \approx a\delta^4(\omega) = a(\pi - \varphi(\omega))^4 \tag{11}$$

with $a = \omega_d / 32$. It is very important to stress that the Taylor expansion of (8) was carried out for $|\delta| \to 0$, but not for small $\Delta\omega = \omega_d - \omega$ as an expansion parameter, owing to the degeneracy condition. In other words, to expand (8) in terms of small $\Delta\omega$ it would require the use of a fractional power series, often called Puiseux series [32], [34], when dealing with a degenerate system. The four steady state solutions following from the approximate result (11) obtained with KVL near the DBE are shown with squares in Fig. 2(a) and are consistent with the exact eigenmode solution and dispersion diagram obtained by solving (4) for $\varphi(\omega)$ with the transfer matrix method near the DBE. However, when $\omega \ll \omega_d$ this first order approximation deviates from the exact calculations as demonstrated in Fig. 2. From (5) and (11) we get the current eigenstates near the DBE

$$\begin{aligned}
I_{n+1} &\approx -e^{-j\delta} I_n, & \text{lower loops} \\
I'_n &\approx j\delta e^{-j\frac{\delta}{2}} I_n, & \text{upper loops}
\end{aligned} \tag{12}$$

The resulting state vector near the DBE for the T/π and π/T configurations in Fig.1 are constructed from the voltage and current of the loops, and omitting details, one gets

$$\begin{aligned}
\Psi_m^{T/\pi}(0) &= \left[(j^{m+1}\tfrac{\delta}{2})^2 \quad 1 \quad (j^{m+1}\tfrac{\delta}{2}) \quad (j^{m+1}\tfrac{\delta}{2})^3\right]^T + O(\delta^4), \\
\Psi_m^{\pi/T}(0) &= \left[(j^{m+1}\tfrac{\delta}{2})^3 \quad (j^{m+1}\tfrac{\delta}{2}) \quad 1 \quad (j^{m+1}\tfrac{\delta}{2})^2\right]^T + O(\delta^4)
\end{aligned} \tag{13}$$

with $m=1,2,3,4$, and they are scaled such that the current (voltage) in upper (lower) ladder is unity in the T/π (π/T) topology, respectively. Near the DBE radian frequency $\omega_d$, the incremental phase angle is given by the 4th root $\delta(\omega) \approx s_m |a|^{1/4} |\omega_d - \omega|^{1/4}$, where we assume $|.|^{1/4}$ to be the principal fourth root, then $s_m = \{1, -1, j, -j\}$ for $\omega < \omega_d$, while $s_m = \{j(1-j)/\sqrt{2}, -j(1-j)/\sqrt{2}, j(1+j)/\sqrt{2}, -j(1+j)/\sqrt{2}\}$, for $\omega > \omega_d$. Complex branches of the dispersion are shown in Fig. 2(b). As such, we have four solutions for $\delta$, except at the DBE frequency where all states coalesce to a single degenerate state with $\delta = 0$, for $\omega = \omega_d$. As mentioned in Section II, for $\omega > \omega_d$ a stop band forms and inhibits signal flow along the double ladder since all $\delta$ solutions are complex-valued.

## IV. RESONANCES IN DOUBLE LADDERS WITH FINITE NUMBER OF CELLS

### A. State Vector and Boundary Conditions

We now consider a lossless circuit with $N$ unit cells (versus infinite cascaded unit cells as in previous sections) as in Fig. 4, whose unit cell is depicted in Fig. 1(c-d), with a signal generator and terminal impedances, as depicted in Fig. 4. Other terminations would not alter the conclusions of this Section. In general, close to (but not exactly at) the DBE, the state vector $\Psi(n)$ at the $n^{\text{th}}$ node in the circuit ($n = 0,1,2,\ldots,N$) is written in terms of the eigenvectors $\Psi_m(0)$ of (2) at location $n = 0$ as

$$\Psi(n) = \sum_{m=1}^{4} c_m e^{-j\varphi_m n} \Psi_m(0) \tag{14}$$

with $m = 1, 2, 3, 4$ and $c_m$ being unknown coefficients which indicate the weights of the four periodic eigenstates excited by the generator. (Note that exactly at the DBE frequency the expansion should be done with generalized eigenvectors since they form complete basis.) The weights $c_m$'s depend on left and right boundary conditions (BCs), i.e., the all terminations. In the case seen in Fig. 4, the upper node at node $n = N$ is terminated with a load impedance $Z_L$, and a generator with voltage $v_g$ and impedance $Z_L$ is located at the upper ladder node $n = 0$. The lower end nodes are shorted in this specific example. These BCs put constrains on the state vectors at $n = 0$ and $n = N$ leading to

$$\begin{aligned}
\Psi(0) &= \left[(v_g - Z_L I_1(0)) \quad 0 \quad I_1(0) \quad I_2(0)\right]^T, \\
\Psi(N) &= \left[Z_L I_1(N) \quad 0 \quad I_1(N) \quad I_2(N)\right]^T
\end{aligned} \tag{15}$$

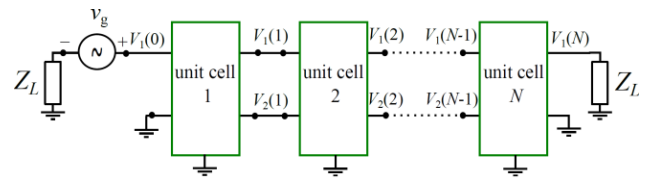

Fig. 4. A double ladder finite size array driven at one end with a voltage generator. The upper nodes are loaded with impedance $Z_L$ at both ends. The lower end nodes are grounded at both ends.

### B. Transfer function and DBE resonance

We define a voltage transfer function as $T_F(\omega) = V_1(N)/V_1(0)$ that is calculated numerically using the transfer matrix method and depicted in Fig. 5 varying as a function of normalized angular frequency near the DBE for the T/π topology in Fig. 1(c), assuming $Z_L = R_L$ to be purely real, and taking $R_L = Z_c = \sqrt{L/C}$ as an example. For clarity we also plot the quantity $|V_1(N)/(v_g/2)|$ which, for the circuit in Fig. 4, is always less than or equal unity. Circuit simulations were also carried out using Keysight ADS and the results in Fig. 5(b) show identical match between the transfer matrix analysis and ADS simulations for the voltage transfer function (and all other circuit quantities, not reported here for brevity).

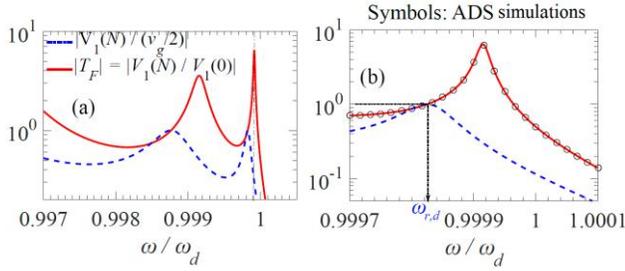

Fig. 5. (a) Transfer function for a finite double ladder with $N=17$ unit cells shown in Fig. 4, with and $R_L=Z_c$, defined as $T_F(\omega)=V_1(N)/V_1(0)$ and we also superimpose the quantity $|V_1(N)/(v_g/2)|$. (b) Zoomed version of (a) around the DBE. The sharpest peak of $|V_1(N)/(v_g/2)|$ is the one occurring at $\omega=\omega_{r,d}$, at which $|V_1(N)|=|V_1(0)|$. Symbols are relevant to ADS circuit simulations.

Several resonances are observed at angular frequencies $\omega_{r,k}$, at which $|V_1(N)|=|V_1(0)|$ or $|V_1(N)|=v_g/2$, for a lossless circuit, and $k$ is the order of the resonance with $k=1,2,3,\ldots,d$, in the close vicinity of the DBE. The closest transmission resonance's frequency to DBE radian frequency $\omega_d$, denoted by $\omega_{r,d}$ (i.e., with subscript $k=d$) is the narrowest and the most significant. We show here how the transmission resonance mode can be analytically calculated using the asymptotic expansion for the eigenstates near the DBE, discussed in Section III. Since the transmission resonance angular frequency $\omega_{r,k}$ is in close vicinity of the DBE, i.e., relation (11) is still satisfied. At the transmission resonance, the input impedance seen to the right at $n=0$ from the generator side is equal to $R_L$ for the lossless and symmetric double ladder circuit in Fig. 4.

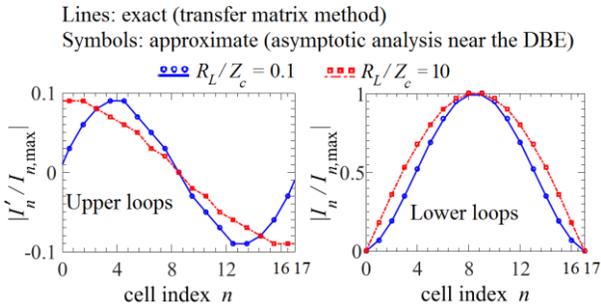

Fig. 6. Normalized loop currents $I'_n$ and $I_n$ at the resonance frequency closest to the DBE, for a $T/\pi$ double ladder of $N=17$ unit cells. Two cases are considered: $R_L/Z_c=0.1$ (blue solid lines) and $R_L/Z_c=10$ (red dashed-dotted lines). Symbols are based on the asymptotic analysis, whereas lines are based on the exact calculations using the transfer matrix method, in very good agreement. For the case with $R_L/Z_c=0.1$ (=10) the resonance normalized frequency $\omega_{r,d}/\omega_d$ is equal to 0.99995 (0.9998) respectively. The current is sampled only at the beginning of each unit cell.

Therefore, the state vectors' quantities, i.e., voltage and current at the two boundaries at $n=0$ and $N$ depicted in Fig. 6 are equal in magnitude but vary in phase thanks to the symmetry of the circuit, and both BCs in (15) are related via $\boldsymbol{\Psi}(N)=e^{-j\theta_{r,k}}\boldsymbol{\Psi}(0)$, where $\theta_{r,k}$ represents the accumulated phase shift across the $N$ cell double ladder. Here we aim also at calculating $\theta_{r,k}$ in terms of $\varphi_{r,k}(\omega_{r,k})$ which is the periodic (real and positive) phase shift of the eigenstate across one unit cell at the transmission resonance radian frequencies. We analytically evaluate the resonance frequency $\omega_{r,k}$ as well as the state vector at any location in the circuit at resonance by carrying out the following steps: Using (14) and the BCs in (15) we construct a system of equations to calculate the $c_m$ coefficients, as a function of frequency. Once the $c_m$ coefficients are obtained as a function of frequency, the state vector at any location in the circuit is readily found from (14). Subsequently, the resonance frequency $\omega_{r,k}$, and phase shift $\theta_{r,k}$ are found by applying the resonance condition $|T_F(\omega_{r,k})|=1$ or equivalently $|V_1(N)|=|V_1(0)|$. In the following we provide analytic expressions of the resonance frequency, transfer function and state vectors.

The possible transmission resonance angular frequencies $\omega_{r,k}$ (or equivalently we calculate the phase shift $\varphi_{r,k}$ at $\omega_{r,k}$ since the asymptotic dispersion relation is known from (11)) near the DBE are obtained by solving transcendental equation

$$\gamma_{r,k}\left(\frac{Z_c}{R_L}\right)^2 \cong N\left[\tanh(\gamma_{r,k})\cot(\gamma_{r,k})-\tan(\gamma_{r,k})\coth(\gamma_{r,k})\right] \quad (16)$$

for $\gamma_{r,k}$, where we have defined $\gamma_{r,k}=N\delta_{r,k}/2$, and $\delta_{r,k}\equiv\varphi(\omega_{r,k})-\pi$ is obtained from (11). It can be easily shown from (16) that for large load resistance, $R_L \gg Z_c$, the resonance frequency shifts (slightly) away from the DBE to lower frequencies, whereas for low load resistance $R_L \ll Z_c$ the resonance frequency tends to approach the DBE frequency. At the resonance angular frequency $\omega_{r,k}$ obtained from (16), the total phase shift $\theta_{r,k}$ of the $N$ double ladder circuit at resonance $\omega_{r,k}$ is obtained from

$$\tan^2\left(\frac{\theta_{r,k}}{2}\right) \cong \frac{\tan(\gamma_{r,k})-\tanh(\gamma_{r,k})}{\cot(\gamma_{r,k})+\coth(\gamma_{r,k})} \quad (17)$$

Currents at any node $n=0,1,2,\ldots,N$ in the double ladder are calculated as the state vector current elements, at frequencies $\omega_{r,k}$ that are close to the DBE as

$$I_1(n) \cong \frac{I_0}{\delta_{r,k}^2}\left[a_2\cos(\xi_n)-a_3\cos(j\xi_n)-a_0\sin(\xi_n)+a_1\sin(j\xi_n)\right],$$

$$I_2(n) \cong I_0\delta_{r,k}\left[a_2\cos(\xi_n)+a_3\cos(j\xi_n)-a_0\sin(\xi_n)-a_1\sin(j\xi_n)\right] \quad (18)$$

where $\xi_n=n\delta_{r,k}-\gamma_{r,k}$, and $I_0=jv_g e^{-j\theta_{r,k}/2}/(4Z_c)$ with

$$a_0=\frac{\cos(\frac{\theta_{r,k}}{2})}{\cos(\gamma_{r,k})}, \qquad a_1=-\frac{\cos(\frac{\theta_{r,k}}{2})}{\cosh(\gamma_{r,k})},$$

$$a_2=-j\frac{\sin(\frac{\theta_{r,k}}{2})}{\sin(\gamma_{r,k})}, \qquad a_3=j\frac{\sin(\frac{\theta_{r,k}}{2})}{\sinh(\gamma_{r,k})} \quad (19)$$

The transfer function of such circuit at any of the resonance frequencies $\omega_{r,k}$ near the DBE is then obtained as

$$T_F(\omega_{r,k}) \cong \frac{I_0 Z_L}{2v_g \delta_{r,k}^2}\big[a_2\cos(\xi_N) - a_3\cos(j\xi_N) +$$
$$-a_0\sin(\xi_N) + a_1\sin(j\xi_N)\big]$$
$$= e^{-j\theta_{r,k}} \qquad (20)$$

which implies that $|T_F(\omega_{r,k})| = 1$. As indicated earlier, the most prominent feature of the circuit is the transmission resonance closest to the DBE whose resonance frequency is denoted by $\omega_{r,d}$, and is also found from solving (16) and by taking the angular frequency $\omega_{r,k}$ that is closest to $\omega_d$. For illustration, in Fig. 6 we show the magnitude of the current in the double ladder made of $N = 17$ unit cells at the DBE-related resonance whose angular frequency is $\omega_{r,d}$, for two cases of load resistance $R_L = 0.1Z_c$ and $R_L = 10Z_c$. Results in Fig. 6 show good agreement with the asymptotic analysis (symbols) in (18) and (19), and the exact one (lines) calculated using the transfer matrix method in Section II. Note that exactly at the DBE a general solution is represented as a composition of *generalized eigenvectors* discussed in Section II, therefore for resonances very close to the DBE, even if we have four independent eigenvectors, the solution still grows since it shows the DBE feature. Furthermore, the weights of those excited eigenstates depend on the load impedance as will be shown next. Such analysis provides insight into the behavior of the circuit, and importantly to analytically show what happens when loads change. Note that one can draw analogous conclusions for higher order ladder circuits operated at the points of degeneracy.

*C. Load resistance effect on transmission phase and resonance frequency*

The current distribution at the DBE resonance whose angular frequency is $\omega_{r,d}$, in Fig. 6 for $T/\pi$ double ladder, shows that most of the energy is concentrated in the lower ladder near the central cells. While $I_n$ is not significantly changed by the load value, $I'_n$ is affected by the large or small load resistance compared to the characteristic impedance $Z_c$. This is related to the weight of the excited state eigenvectors, namely $c_m$ in (14), required to match the BCs at the two terminations at $n = 0$ and $n = N$. It can be inferred that when the load resistance is low compared to the characteristic impedance, i.e., $R_L \ll Z_c$, the coefficients $c_m$'s of the two eigenstates vector with *complex* phase shift $\varphi$ are negligible in magnitude, compared to the same for the two eigenstates vectors with purely *real* phase shift near the DBE. On the contrary, for $R_L \gg Z_c$, all the eigenstates are excited with comparable weights $c_m$. This is observed from the numerical solution of (17) for the $T/\pi$ double ladder in Fig. 1. The opposite trend is observed for the $\pi/T$ double ladder, not shown here for brevity. Moreover, the terminal impedance has an impact on the transmission phase as indicated in (16) and (17). To demonstrate that effect, we show in Figs. 7(a) and (b) the normalized DBE resonance frequency $\omega_{r,d}/\omega_d$ and the total phase shift $\theta_{r,k}/\pi$, respectively, varying as function of the normalized load resistance $R_L/Z_c$ for a double ladder of $N = 16$ and $N = 17$ unit cells calculated using the transfer matrix method developed in Section II. As seen in Fig. 7(a), the resonance frequency slightly changes (scale is zoomed to show the insignificant variation) whereas the change in phase shift in Fig. 7(b) is significant when the load varies from low to high, though it varies slowly. This can be also inferred from solving (16) and (17). The total phase shift at resonance $\theta_{r,k}$ varies from $\sim 0$ rad for low impedance loads to $\sim \pi$ for high impedance loads, for odd number of cells (vice versa for even number of cells). However, the resonance frequency always shifts slightly to lower frequencies than the DBE frequency as load impedance increases. We also use Keysight ADS for simulating the circuit behavior and the results in Fig. 7 show identical match between the transfer matrix analysis and ADS simulations.

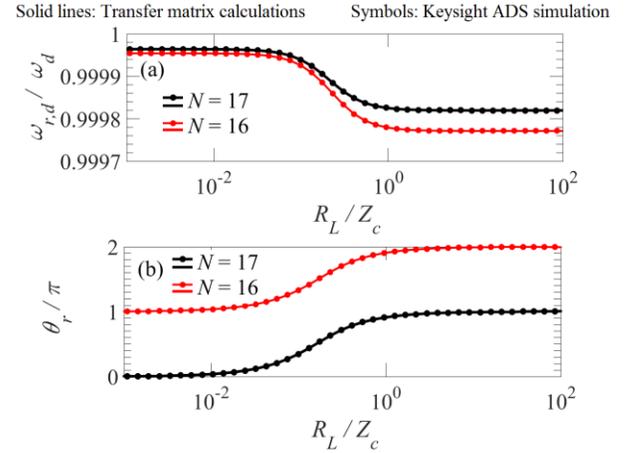

Fig. 7. (a) Small variation of the resonance frequency of the T/π double ladder circuit closest to DBE varying as a function of the load impedance. (b) The corresponding transmission phase $\theta_{r,d}$. Cases with for $N = 16$ and 17 cells are shown. Identical results are obtained using transfer matrix as well as ADS.

*D. Load impedance effect on total Quality factor*

The loaded or total quality factor, denoted by $Q_{tot}$, of a reactive circuit [pp. 302, 37] such as the one in Fig. 4 is defined as

$$Q_{tot} = \frac{\omega_r[W_e + W_m]}{P_l} \qquad (21)$$

where $W_e$, $W_m$, and $P_l$ are the total time-average stored electric energy, stored magnetic energy, and power loss, respectively, given by

$$W_e = \frac{1}{2}C\sum_{p=1}^{2N}|V_{Cp}|^2, \quad W_m = \frac{1}{4}L\sum_{q=1}^{4N}|I_{Lq}|^2 \qquad (22)$$

and $P_l = \frac{1}{2}R_L\big(|I_1(0)|^2 + |I_1(N)|^2\big)$, all at $\omega_{r,k}$. For simplicity here we consider only circuits with no internal losses, and losses occur only at the termination load. Here $V_{Cp}$ and $I_{Lq}$ are the voltage across the $p^{th}$ capacitor and the current in the $q^{th}$ inductors in the double ladder circuit, respectively, that are excited by the generator, and are easily related to the node voltages and currents. For the symmetric T/$\pi$ double ladder circuit we calculate the loaded $Q_{tot}$ factor of the resonance at $\omega_{r,d}$ and show it in Fig. 8 versus load impedance. Here we use the transfer matrix method to obtain the state vector voltage and current at any node in the circuit, thereupon we use (21) to numerically calculate the loaded $Q_{tot}$ factor. Indeed, the loaded $Q_{tot}$ has a minimum at the same value of $R_L$ where the total phase shift $\theta_{r,d}$, in Fig. 7(b), has the steepest slope. In the Appendix we provide an analytic expression for the loaded $Q_{tot}$ factor using the asymptotic analysis developed in this paper for large $N$. It can be inferred that $Q_{tot} \approx Q_{tot,min} \cosh^2(R_L/Z_c - \zeta)$ where $\zeta$ is a fitting constant (see the Appendix). Indeed, the loaded $Q_{tot}$ varying as a function of $R_L$ has a minimum value denoted by $Q_{tot,min}$ that occurs at a certain load impedance $R_{L,min}$, for a fixed ladder size. Therefore, we notice the important property that $Q_{tot}$ of the resonant state of the circuit close to the DBE exhibits slight dependence on the load value $R_L$, since the variations in $\theta_{r,d}$ with loading are smooth as seen in Fig. 7(b), contrary to what happens in single ladders which is discussed in the next section. It is worth mentioning that it is important to study the behavior of $Q_{tot,min}$ occurring at $R_{L,min}$, because such load $R_{L,min}$ corresponds to the maximum power delivered to the load in oscillators [36]; which also would be the preferable load for minimizing phase noise [the reader is referred to [36] for preliminary investigation of the double ladder oscillators]. Note that such unusual behavior of the $Q$ factor is often referred to as *giant resonance* [4], [5], [10], [37]. Here we demonstrate rigorously the unusual behavior of the DBE resonance of double ladders in terms of stability of resonance frequency against loading and the scaling of the loaded Q factor.

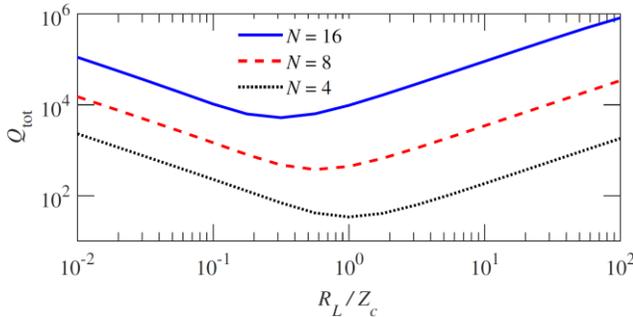

Fig. 8. Loaded Quality factor versus load resistance at the DBE resonance $\omega_{r,d}$ of a circuit with a finite number of unit cells ($N$ =4, 8, 16). The quality factor never decreases to low values for any loading. The minimum quality factor for each given $N$-length, $Q_{tot,min}$, slightly shifts to lower load resistances when increasing the double ladder size $N$.

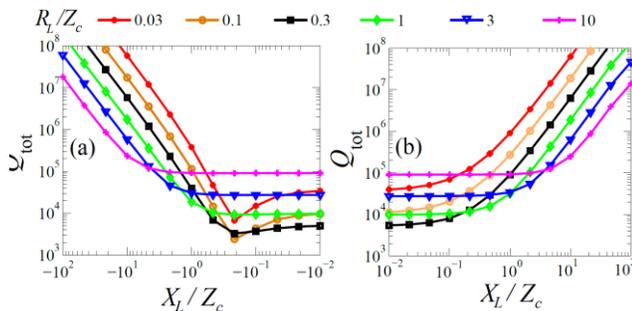

Fig. 9. Quality factor versus normalized load reactance (capacitive loads with $X_L < 0$ on the left, inductive loads with $X_L > 0$ on the right) at the DBE resonance $\omega_{r,d}$ for various values of load resistance $R_L/Z_c$, and $N$ = 16.

We also report how $Q_{tot}$ changes with a reactive load, in which the load impedance is given as $Z_L = R_L + jX_L$. The reactive component of the load contributes to increasing $Q_{tot}$ for a lossless double ladder, as can be seen in Figs. 9(a) and (b), for both capacitive and inductive loads. Remarkably, the $Q_{tot}$ is stable (independent of the reactance value) within the range $-1 < X_L/Z_c < 1$ especially for $R_L/Z_c > 1$.

## V. SINGLE LADDER VERSUS DOUBLE LADDER CIRCUIT COMPARISON

We compare transmission phase characteristics across ladders of finite size as a function of the load resistance $R_L$ for the two cases of single and double ladders, supporting an RBE and a DBE, respectively. A single ladder is constructed by cascading unit cells and terminating both ends by $Z_L$, with both conventional T and $\pi$ topologies, in which the inductors and capacitors $L$ and $C$ have the same values as in their double ladder counterpart (Fig. 1). Accordingly, these periodic single ladder circuits develop an RBE, occurring at an angular frequency that coincides with the DBE angular frequency $\omega_d = 1/\sqrt{LC}$ of the double ladder. For the sake of assessment, we compare single and double ladder of the same size, i.e., same number $N$ of unit cells, in addition to having the same termination impedances for the double and single ladders. We compare four topologies, two pertaining to *double ladders* ($T/\pi$ and $\pi/T$ topologies as in Fig. 1) and two pertaining to *single ladders* (T and $\pi$ topologies as in Fig. 1). The total phase shifts across the single/double ladders, namely $\theta_{r,d}$, are defined as the phase of the transfer function $T_F(\omega_{r,d}) \equiv V_1(N)/V_1(0)$ calculated at their respective transmission resonance angular frequencies that are the closest to $\omega_d = 1/\sqrt{LC}$ in each of the four cases (recall that some properties of band edge resonance in RBE periodic structures are discussed in details in [4], [38]). In Fig. 10 we show the transition of the total phase shift $\theta_{r,d}$ varying as a function of the load resistance $R_L$ for four topologies for the resonance closest to $\omega_d$. It is important to observe that the phase shift as a function of load is qualitatively different for the single ladder and double ladder, which is a non-intuitive feature for these circuits.

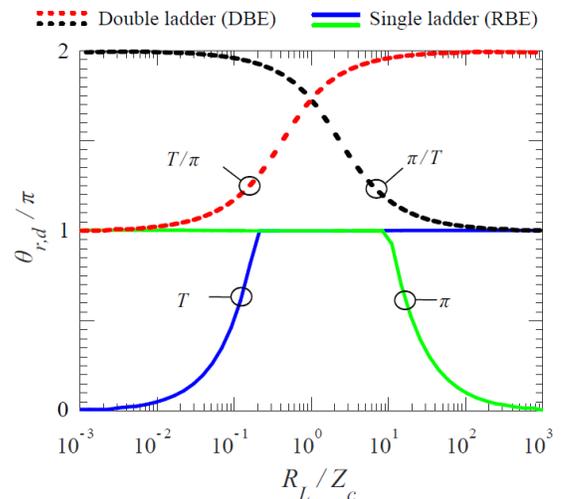

Fig. 10. Total transmission phase shift $\theta_{r,d}$ across the circuit in Fig. 4 made of $N$=8 unit cells versus load resistance for *double* and *single* ladder topologies. Notice the abrupt phase transition in single ladder's phase shift versus load impedance, compared to the double ladder.

The single ladder T-topology's phase transitions abruptly from zero phase shift at low load resistance to $\pi$ at some high load resistance. The transition is abrupt because the single ladder has a high Q transmission resonance at the RBE frequency for low-impedance terminations, that appears (disappears) for high impedance loads in the $\pi$ (T) configuration [23]. In other words, in an $N$ cell single ladder, the number of transmission resonances at very low and high load impedance is $N+1$ and $N$ respectively for the T topology and vice versa for the $\pi$ topology. This may render the single ladder circuit's resonance unstable near this transition (a known phenomenon related to *mode jumping* [23] in single ladder multimode oscillators). In contrast, the T/$\pi$ double ladder topology (with the unit cell shown in Fig. 1(c)) transitions smoothly from $\pi$ at low load resistance to $2\pi$ at high load resistance. The transition is smooth because the DBE resonance makes the transition by maintaining the resonant mode at a frequency lower than the DBE frequency and therefore it is stable. Thus the number of resonance modes close to the DBE is conserved; and the DBE resonance has a stable resonance frequency not prone to load variation contrary to what occurs for single ladders. This indicates that the total $Q_{tot}$ factor of the single ladder circuit can be quite different for some specific variations of load impedances and its resonance frequency is not stable. Obviously, for extremely high/low load impedances, loading effects on the $Q_{tot}$ are insignificant.

Now we compare in Fig. 11 the performance of a finite double ladder circuit to that of a finite single ladder in terms of $Q_{tot}$ when the load resistance varies. Note that despite the total *inductance* and *capacitance* of the single ladder unit cell is half that of the double ladder unit cell (Fig. 1), only the lower ladder in the DBE configuration stores most of the energy, in the sense that the upper ladder nodes are essentially shorted (RF ground) as well as the middle node in T/$\pi$ lower ladder unit cell (see the current distribution in Fig. 6). Therefore, the total number of elements that store energy is effectively the same in both configurations. The $Q_{tot}$ factor of both circuits has a minimum when varying the load value, denoted as $Q_{tot,min}$ corresponding to a transition of the transmission phase shift as discussed earlier. Note that the value of the load resistance at which $Q_{tot} = Q_{tot,min}$ differs from $Z_c$ as $N$ increases, as discussed in Section IV. Observe also that the double ladder has higher $Q_{tot,min}$ for larger $N$ (e.g., $N$ = 8, 16) than the single ladder. Such load impedance at which $Q_{tot} = Q_{tot,min}$ diverges from $Z_c$ more rapidly in the single ladder compared to the same in the double ladder, indicating that double ladders tend to maintain its resonance frequency and quality factor regardless of the load.

As shown in Fig. 12 the minimum quality factor $Q_{tot,min}$ increases with increasing number of cells for both single and double ladders. However, for long ladders (i.e., for $N > 5$) the double ladder has significantly higher $Q_{tot,min}$ than single ladders Remarkably, the loaded quality factor for a double ladder scales as $Q_{tot,min} \propto N^5$ while the same for the single ladder scales as $Q_{tot,min} \propto N^3$. For small sized ladders ($N < 5$) the single ladder may have comparable loaded quality factor since the DBE feature rises for sufficiently large $N$.

For these reasons, double ladder oscillators are the immediate application of such resonance circuit that provides for low threshold as well as stable oscillator frequency to drive a 50 Ω impedance load, without the need of external current mode logic (CML) buffers to drive such load.

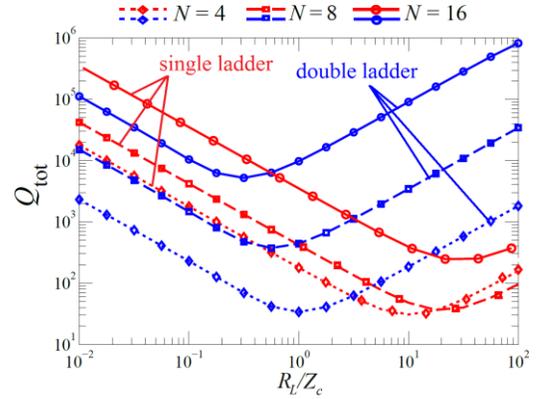

Fig. 11. Quality factor versus normalized load resistance $R_L/Z_c$ for both single and double ladders. Notice how it never reaches very low values, for any loading, especially for circuits with larger number of unit cells. Double ladders show smaller variation of quality factor.

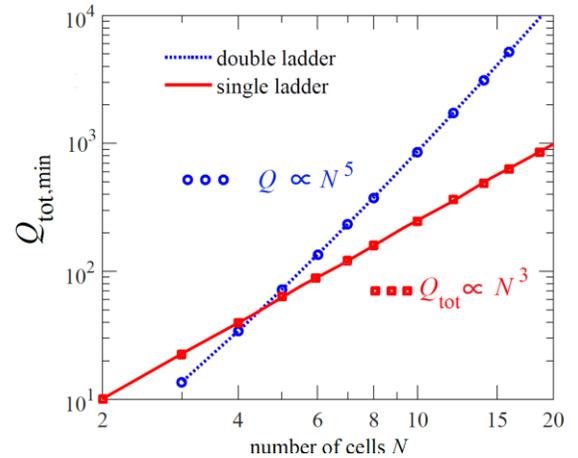

Fig. 12. Minimum quality factor $Q_{tot,min}$ vs number of unit cells $N$ (in log-log scale) for double ladders (squares) and for single ladders (circles). Curves are fitted with $N^5$ and $N^3$ trends respectively (lines). For circuits having more than 5 unit cells, the double ladder has always larger loaded quality factor and has the special growing trend as $N^5$.

## VI. CONCLUSION

We have presented for the first time a comprehensive theoretical formulation that explains the physical behavior and the loading properties of double ladder periodic circuits with a fourth order degeneracy. We have demonstrated that a periodic circuit whose unit cell is made of only five lumped elements exhibits a degenerate band edge in the phase-frequency dispersion relation; and we have shown analytically the eigenstates (voltage/current) behavior of such periodic circuit near the DBE. We have also analytically derived the transfer function, resonance frequency and the total quality factor near the DBE in such double ladders. The analytical theory developed here can be utilized to analyze and design other DBE structures.

The double ladder circuit has several advantages over an equivalent single ladder of the same size, in terms of stability of the resonance frequency and quality factor against loading effects. A double ladder exhibits unusual scaling of the loaded $Q_{tot}$ as $N^5$, where $N$ is the number of unit cells, versus that of the single ladder that scales as $N^3$. The finite double ladder is less sensitive in several respects to variation in the load resistance and has a higher $Q_{tot}$ than the single ladder for somewhat large $N$. Moreover, a double ladder may provide for a low-threshold resonance conditions for oscillators, with the property that the threshold as well as oscillation frequency is weakly dependent on the load as it will be shown in future. In particular, a double ladder would operate with a *single frequency* because the resonance near the DBE has the highest quality factor $Q_{tot}$ for a certain load, therefore, due to nonlinearity and saturation effects discussed in [36], the oscillation frequency will more or less coincide with the resonance at the DBE. In addition, the oscillation *frequency is independent of loading* contrary to conventional LC tank oscillator or even single ladder counterpart. Other applications of such circuit may include pulse forming delay lines, pulse compressors, filter, and distributed amplifiers.

## APPENDIX: ANALYTICAL EXPRESSIONS FOR THE TRANSFER MATRIX AND THE QUALITY FACTOR

Although the results and conclusions reported in this paper are independent of the values of $L$, and $C$, yet we have used $L = 45$ μH and $C = 65$ pF that provide for a DBE condition at 100.23 MHz throughout this paper. These values of L and C can be obtained with commercially available discrete off-the-shelf components. We stress that the main circuit parameters are the DBE frequency and the characteristic impedance; not the precise value of L and C. Yet, values and frequency can be properly scaled, and other circuit topologies can be devised as well. The transfer matrix of the unit cell in Fig. 1 is calculated by multiplying the 4×4 matrices of each individual element. The calculation is cumbersome, but after simplification the transfer matrix reads,

$$\underline{\mathbf{T}} = \begin{bmatrix} 1+\Gamma & -\Gamma(1+\Gamma/2) & j\omega L(2+\Gamma) & -j\omega L\Gamma \\ -\Gamma & 1+2\Gamma+\Gamma^2/2 & -j\omega L\Gamma & j\omega L(2+\Gamma) \\ 2j\omega C & -j\omega C(2+\Gamma) & 1+\Gamma & -\Gamma \\ -2j\omega C(1-\Gamma) & j\omega C(4+\Gamma)(1+\Gamma/2) & -\Gamma(1+\Gamma/2) & 1+\Gamma(2+\Gamma/2) \end{bmatrix} \quad (A1)$$

where $\Gamma = -2(\omega/\omega_d)^2$. Analytic expressions for the state-vector solution at resonances $\omega_{r,k}$ is given in [39].

To calculate the quality factor of the double ladder resonator with $N$ unit cells at the resonance frequency $\omega_{r,d}$, we conveniently assume that all the energy is stored in the lower ladder branch therefore the capacitor and inductors in the upper ladder and coupling branches store negligible energy. That is compliant with the characteristics of the voltage distribution seen in Fig. 6 and with having proven that $I'_n = 0$ exactly at the DBE. Using the currents of the nodes as analytically expressed in (21), we write the loaded $Q_{tot}$ factor of the DBE resonant mode whose frequency is $\omega_{r,d}$ as

$$Q_{tot} \cong \frac{Q_0}{\left((\omega_{r,d}/\omega_d)^2 - 1\right)} \left(1 + \left|\frac{I_2(1)}{I_2(0)}\right|^2 + \cdots + \left|\frac{I_2(\frac{N}{2})}{I_2(0)}\right|^2\right) \quad (A2)$$

where $Q_0 = \omega_d L / Z_c$. Note the term $((\omega_{r,d}/\omega_d)^2 - 1)$ in the denominator that is responsible for the large enhancement of $Q_{tot}$ near the DBE, since the resonance $\omega_{r,d}$ very rapidly approaches $\omega_d$ as $N$ increases. This term $((\omega_{r,d}/\omega_d)^2 - 1)$ is also strongly dependent on the load $R_L$ as discussed next. The second term in (A2) simply represents a sum of the magnitude of the normalized current in the circuit. We analyze two cases:

*i)* Variation of $Q_{tot}$ as a function of $N$ for constant load resistance $R_L$. In that case, the term $((\omega_{r,d}/\omega_d)^2 - 1)$ is proportional to $1/N^4$ [4], [29]. The contribution of the current summation term $1 + 2|I_2(1)/I_2(0)|^2 + \cdots + |I_2(N/2)/I_2(0)|^2$ to $Q_{tot}$ depends on the load, but for the specific circuit in Fig. 4 it is numerically shown that it is proportional to $N$, hence the $Q_{tot}$ of the circuit in Fig. 4 is proportional to $N^5$, regardless of the specific value of the load $R_L$.

*ii)* Variation of $Q_{tot}$ as a function of load resistance $R_L$ for constant $N$. The load resistance also affects the term $((\omega_{r,d}/\omega_d)^2 - 1)$ in the way described earlier in (19). It can be seen in Fig. 7(a) that the resonance frequency is varying as a function of the load, in an asymptotic fashion as $\omega_{r,d} \propto \tanh(R_L/Z_c - \zeta)$ where $\zeta$ is a fitting constant, as also deduced from (19) and (20), and in Fig. 8 we have $\zeta \approx 0.3$ for the specific circuit under analysis. Therefore $((\omega_{r,d}/\omega_d)^2 - 1) \propto \text{sech}^2(R_L/Z_c - \zeta)$. Finally, the behavior of $Q_{tot}$ versus load resistance $R_L$ can be expressed as $Q_{tot} \propto \cosh^2(R_L/Z_c - \zeta)$ which confirms that the loaded $Q_{tot}$ factor has a minimum versus load impedance $R_L$, at the location where the total phase shift changes, which happens close to the condition $R_L \approx \zeta Z_c$.